\documentclass{elsart}

\usepackage{graphicx}
\usepackage{epsfig}
\usepackage{amsmath}
\usepackage{url}

\newcommand{\tens}[1]{\mathbf{#1}} 
\newcommand{\pd}{\partial}

\DeclareMathOperator{\De}{De}

\newcommand{\ttt}{\tens{T}}

\newcommand{\tyy}{T_{yy}}

\begin{document}

  \begin{frontmatter}

    \title{Stress singularities and the formation of birefringent strands in stagnation flows of
           dilute polymer solutions}
    \author[Leiden]{Paul Becherer\corauthref{cor1}},
      \ead{becherer@lorentz.leidenuniv.nl}
    \author[Leiden]{Wim van Saarloos} and
    \author[Edinburgh]{Alexander N. Morozov}
    \corauth[cor1]{Corresponding author. Tel.: +31 71 527 5517;
                                          fax: +31 71 527 5511}
    \address[Leiden]{Instituut-Lorentz for Theoretical Physics,
                     Universiteit Leiden,
                     Postbus 9506,
                     NL-2300 RA Leiden,
                     The Netherlands
                     }
    \address[Edinburgh]{School of Physics,
                        University of Edinburgh,
                        JCMB,
                        King's Buildings,
                        Mayfield Road,
                        Edinburgh EH9 3JZ,
                        United Kingdom
                      }
  \begin{abstract}
    We consider stagnation point flow away from a wall for
    creeping flow of dilute polymer solutions. For a simplified
    flow geometry, we
    explicitly show that a narrow region of strong polymer extension
    (a birefringent strand) forms downstream of the stagnation point
    in the UCM model and extensions, like the FENE-P model. These
    strands are associated with the existence of an essential
    singularity in the stresses, which is induced by the fact that the
    stagnation point makes the convective term in the constitutive
    equation into a singular point. We argue that the mechanism is
    quite general, so that all flows that have a separatrix going away
    from the stagnation point exhibit some singular behaviour.
    These findings are the counterpart for wall stagnation points of
    the recently discovered singular behaviour in purely elongational
    flows: the underlying mechanism is the same while the different
    nature of the singular stress behaviour reflects the different
    form of the velocity expansion close to the stagnation point. 
  \end{abstract}

  \end{frontmatter}

\section{Introduction}

  Extensional flows of polymer solutions and melts occur in many
  industrial polymer processing operations, and hence such flows have
  been studied for decades \cite{bird1,bird2}. Recently however,
  interest in extensional flows was renewed by observations of steady
  and unstable continuous flow in microfluidic
  devices~\cite{xi2007,arratia2006,poole2007}, and it was
  realized only recently that extensional flows are prone to the
  formation of singularities and non-analytic structures in the stress
  fields \cite{renardy2006,thomases2006,becherer2008}. Depending on
  the Deborah number and the model used, these stress singularities
  may take various forms. For purely extensional flow in continuum
  models that describe infinitely extensible polymer chains (such as
  the upper convected Maxwell model (UCM) and the Oldroyd-B
  model~\cite{bird1,bird2,larson}) the stresses can have power
  law spatial behaviour with a finite limit at the centre line, or
  they even have power law divergencies. For models that are
  based on finitely extensible chains, divergencies are cut off
  at some scale, but singular behaviour of the stress \emph{gradients}
  may persist~\cite{renardy2006,thomases2006,becherer2008}.
 Such singular behaviour may have important implications for numerical
  simulations of extensional flows, since it leads to structures with
  a very small length scale. Indeed it is known that for many such
  flows, numerical schemes break down at only moderate flow rates
  (Deborah numbers of order unity).
  
 The question quite naturally comes up whether singular behaviour
 near special points is the rule rather than the exception. We argue
 in this Communication that the latter is the case and demonstrate
 this for a simplified case where all calculations can be done
 analytically, so that the emergence of the singular behaviour can be
 followed explicitly.
  
  The reason to expect singular behaviour near special points where the
  velocity vanishes --- even though the geometry is not 
  singular\footnote{Of course, at sharp corners where the flow field 
  itself is singular, this singular behaviour carries over to the 
  stresses.} --- is actually very simple. For steady flow, the only 
  derivative terms of the stress $\ttt$ in UCM-type constitutive 
  equations come from the convective term $(\mathbf{v}\cdot\mathbf{\nabla}) 
  \ttt$. The points where $\mathbf{v}$ vanishes --- the stagnation point
  in elongational flow or in the wall stagnation point flow considered
  here --- thus translate into a singular point~\cite{benderorszag1978}
  of the partial differential equation obtained from the constitutive
  equation for the stress. Close to the singular point, the lowest
  order terms in the expansion of $\mathbf{v}$ are often fixed by simple
  symmetry considerations and boundary conditions, if applicable. So
  the nature of the dominant singularity at the singular point is
  generally fixed independent of the precise details of the model.
  Further away from the singularity, the behaviour will typically 
  depend on the details of the flow profile. All these features are
  well illustrated by the analysis below. As stated, we focus on a
  simple case where the calculations can all be done analytically,
  but the scenario holds generally for complex more realistic flows
  and we suspect this mechanism of advection to be at the origin of
  the formation of birefringent strands.

\begin{figure}
    \setlength{\tabcolsep}{5mm}
    \begin{center}
      \vspace{3cm}
      \begin{tabular}{cc}
        \includegraphics[width=0.2\textwidth]{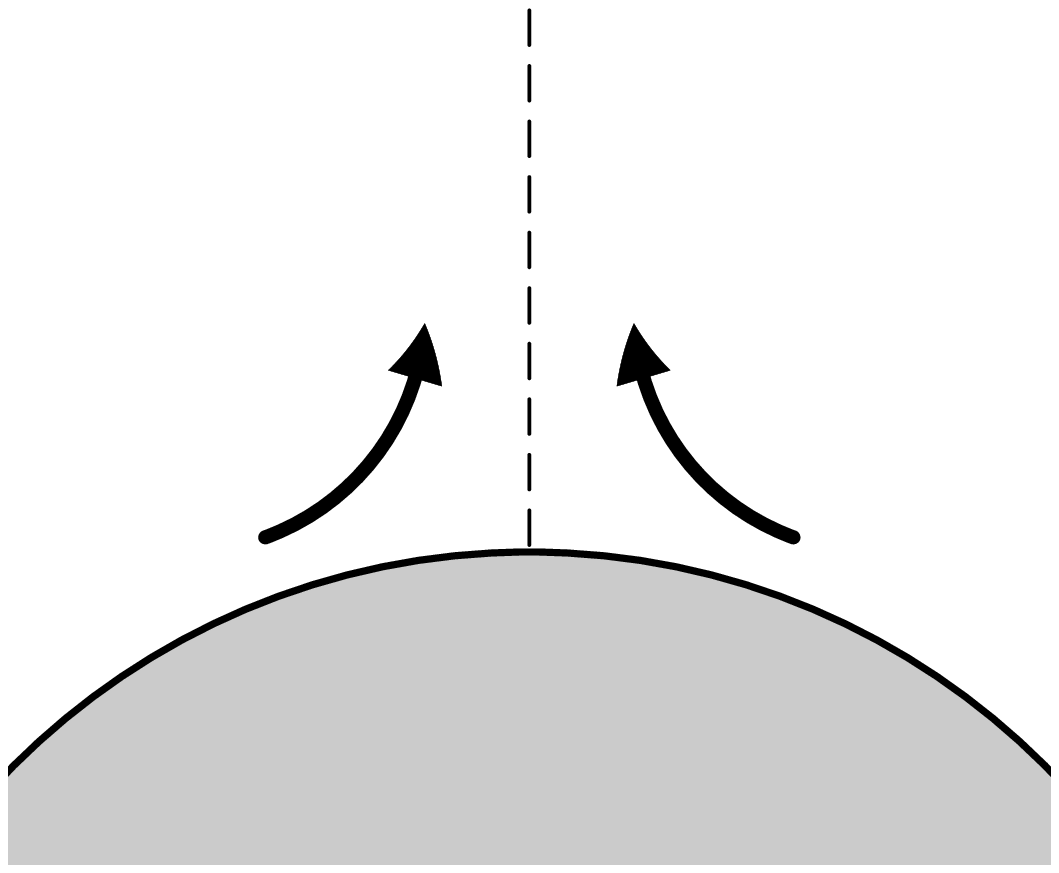} &
        \includegraphics[width=0.2\textwidth]{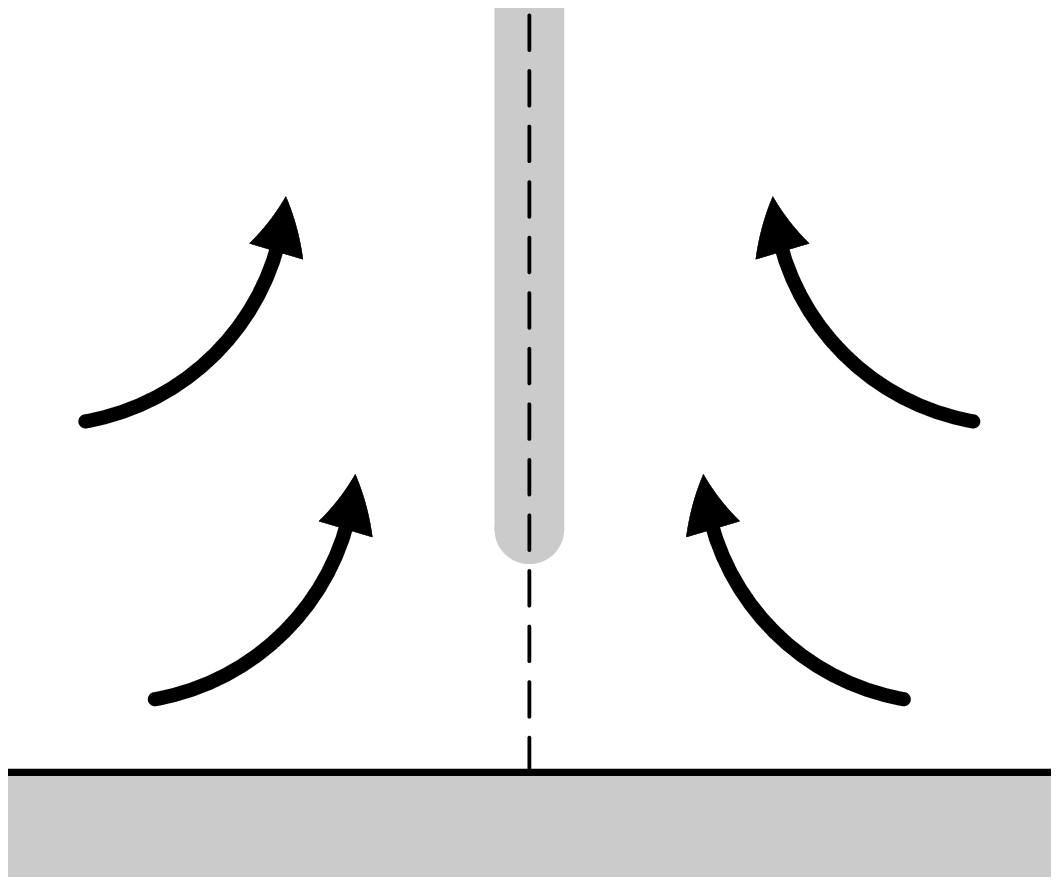} \\
        (a) & (b) 
      \end{tabular}
    \end{center}
    \caption{\label{fig:wall_flow}Stagnation flow (a) in a wake, (b)
    approximated by a flow near a flat wall. In (b) the formation of a
    birefringent strand is qualitatively indicated by the shaded area.}
  \end{figure}

  We focus on wall stagnation point flows where the flow is away from
  the wall. Examples of this are flows in the wake of a falling sphere
  or of a fixed cylinder, as shown schematically
  in Fig.~\ref{fig:wall_flow}~(a). In particular, the flow past a fixed
  cylinder or sphere in a channel has become a benchmark problem for
  numerical modelling of viscoelastic constitutive
  equations~\cite{wapperom2005,fan2005,hulsen2005}. It is known that
  in such flows a narrow region of high polymer extension may form, a
  so-called \emph{birefringent strand}~\cite{cressely1980,harlen1990b}.
  This region starts at a small but finite distance downstream from the
  stagnation point, as indicated schematically in
  Fig.~\ref{fig:wall_flow}~(b), where a flow near a flat wall is
  depicted.
  
  In this work, we consider a strictly two-dimensional version of this
  flow, with a simplified, fixed velocity field obeying the basic
  symmetry of a stagnation point at a wall (cf.~\cite{wapperom2005}).
  We analyse this case in detail for the UCM
  model~\cite{bird1,bird2,larson} but also discuss in the end the
  qualitative changes that occur for a FENE-P model.
 
  Unlike the case of steady purely extensional flow, which was analysed
  previously~\cite{renardy2006,thomases2006,becherer2008}, the
  extension of the polymers does
  not diverge for any extension rate. We find that a thin birefringent
  strand forms, with a singularity at its centre. As argued above,
  notwithstanding the simplifications we make in obtaining this result,
  we believe that the analysis makes it clear how singular
  behaviour emerges in general.
    
  Unfortunately, our results cannot immediately be compared
  quantitatively with experiments or numerical computations on
  realistic cases like flow past a cylinder~\cite{fan2005,hulsen2005}.
  First, one should keep in mind that in such situations, there
  may be two sources of (near) singular behaviour: besides the one
  we analysed here, dominated by the symmetry and boundary conditions
  of the velocity field near the wall stagnation point, in viscoelastic
  flow past a cylinder large stress fields are already built up at
  the sides of the cylinder, where the flow is mostly along the
  cylinder. These shear stresses are advected toward the rear
  stagnation point. This effect is clearly not present in the
  simplified geometry that we consider. Second, our analysis is
  based on taking a \emph{fixed} velocity field obeying the basic
  symmetry, and we show how this leads to an essential singularity
  in the stress. In reality, of course, the velocity and stresses are
  coupled indirectly through the momentum balance equation. Through
  this coupling, the velocity field will also be affected in the region
  of large stress gradients near the stress singularity. Since the
  symmetry and expansion of the velocity near the stagnation point
  cannot change in lowest order, we expect that there is an
  intermediate flow regime where the basic structure of the singularity
  is not changed dramatically. This assumption is further supported by
  recent simulations of a two-dimensional cross-slot flow by
  Poole et al.~\cite{poole2007}. There, the velocity profiles remained
  smooth even when the flow changed its symmetry (the new type of
  purely elastic instability discovered by Arratia et
  al.~\cite{arratia2006}), while stresses exhibit the typical singular
  structure similar to the one discussed
  in~\cite{renardy2006,becherer2008}. At the same time, numerical
  studies suggest
  that at sufficiently large flow rates, this nonlinear coupling can
  become so strong that there may be no steady state flow solution
  past a cylinder for Deborah numbers of order
  unity~\cite{fan2005,bajaj2008}. The coupling and
  this effect are, unfortunately, beyond the present approximation.

  The layout of this paper is as follows. In section 2 we introduce
  the flow geometry and the models, and we briefly recapture
  similarity solutions for UCM found by other
  authors~\cite{phanthien1983,oztekin1997}. We calculate
  analogous solutions for a simplified version of this flow, where
  we fix the velocity field, for UCM and FENE-P. In section 3 we
  consider more realistic boundary conditions, and we solve the
  constitutive equations analytically. In section 4 we consider the
  resulting stress field (extension field) in more detail, showing
  that we find a narrow region of high polymer extension, with a
  non-analytic stress profile at the centre of the strand. We then
  discuss these results in the light of more realistic flow profiles,
  and we conclude by discussing the relevance of these results for
  computational and experimental work.

\section{Simplified stagnation flow of a UCM fluid}
  We consider incompressible planar stagnation flow of a UCM fluid
  without inertia (creeping flow). The UCM constitutive equation
  for steady flow is~\cite{bird1}
  \begin{equation}
    \ttt
      + \lambda \left[ 
        (\mathbf{v}\cdot\mathbf{\nabla}) \ttt
      - (\mathbf{\nabla} \mathbf{v})^T \cdot \ttt
      - \ttt\cdot(\mathbf{\nabla} \mathbf{v})
      \right]
    = \eta \left(
        \mathbf{\nabla}\mathbf{v} + (\mathbf{\nabla}\mathbf{v})^T 
      \right),
      \label{eqn:const_ucm}
  \end{equation}
  where $\lambda$ is the relaxation time of the fluid and $\eta$ is
  the Newtonian viscosity. The momentum balance for creeping flow is
  \begin{equation}
    \mathbf{\nabla}\cdot\ttt - \mathbf{\nabla} p = 0,
    \label{eqn:ns}
  \end{equation}
  where $p$ is the pressure. Incompressibility is given by
  \begin{equation}
    \mathbf{\nabla} \cdot \mathbf{v} = 0.
  \end{equation}

\begin{figure}[b]
\begin{center}
  \includegraphics[width=0.8\textwidth]{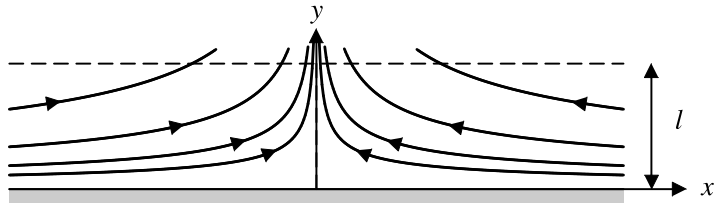}
  \caption{\label{fig:flow_geom} Planar stagnation flow.}
  \end{center}
\end{figure}

  The planar stagnation flow geometry we consider is depicted in
  Fig.~\ref{fig:flow_geom}. We take the vertical direction as bounded,
  with length $\ell$. Because of the solid wall, the boundary condition
  at the wall ($y = 0$) is $\mathbf{v} = 0$. At $y = \ell$, we impose
  $v_y = V$, with $V >0$. For the velocity field, a similarity solution
  then exists, which is of the form~\cite{phanthien1983,oztekin1997}
  \begin{equation}
    v_x = -x \psi'(y) \quad \textrm{and} \quad v_y = \psi(y),
    \label{eqn:simsol}
  \end{equation}
  where the boundary conditions imply $\psi(0) = 0$ and $\psi'(0) = 0$.
  Inspired by this solution, we take a \emph{fixed} velocity field
  that satisfies the boundary conditions and that would correspond
  to the lowest-order approximation near the wall:
  \begin{equation}
    v_x = -2 (V / \ell^2) x y  \quad \textrm{and}
    \quad v_y = (V/\ell^2) y^2.  \label{eq:veq}
  \end{equation}
  Note that these terms are the lowest order analytic terms in an
  expansion in $x$ and $y$ away from a symmetric stagnation point at
  the wall. This is why our analysis illustrates  the emergence of
  singular behaviour in more general cases as well. 
  
  Let us now insert this velocity field into the constitutive
  equation. The resulting stress field will not in general satisfy
  momentum balance, but it does yield a valid solution for
  Newtonian creeping flow. This is sometimes referred to as
  ``Newtonian kinematics''; in the Oldroyd-B extension of the UCM
  model, this would be a reasonable approximation in the dilute
  limit, $\beta \ll 1$, in which the polymer stresses do not
  influence the flow~\cite{oztekin1997,bajaj2008}.
  
  We can rescale the quantities appearing in the equations:
  length is scaled with $\ell$, velocity with $V$, and stress with
  $\eta V / \ell$. The velocity field becomes
  \begin{equation}
    v_x = -2 x y  \quad \textrm{and} \quad v_y = y^2,
    \label{eqn:velocity}
  \end{equation}
  with $0 \leq y \leq 1$, and for the constitutive
  equation we obtain a dimensionless form
  \begin{equation}
    \ttt
      + \De \left[ 
        (\mathbf{v}\cdot\mathbf{\nabla}) \ttt
      - (\mathbf{\nabla} \mathbf{v})^T \cdot \ttt
      - \ttt\cdot(\mathbf{\nabla} \mathbf{v})
      \right]
    = \mathbf{\nabla}\mathbf{v} + (\mathbf{\nabla}\mathbf{v})^T .
      \label{eqn:const_ucm_scaled}
  \end{equation}
  Here we introduced the Deborah number\footnote{One should keep
  in mind that our Deborah number cannot directly be compared with
  the one used in studies of flow past a cylinder.}
  \begin{equation}
    \De = \lambda V / \ell.
  \end{equation}
  We can then insert the rescaled velocity into the constitutive
  equation and solve for the stresses.
  We obtain equations for the components of the stress tensor,
  and we observe that the equation for the $yy$ component decouples
  from the other equations, because $\pd v_y / \pd x$ is identically
  zero for this velocity field. The equation for the $yy$ component
  becomes
  \begin{equation}
    \tyy + \De \left[ -2 x y \frac{\pd \tyy}{\pd x}
                      + y^2  \frac{\pd \tyy}{\pd y}
                      - 4 y \tyy 
               \right] = 4 y.
  \label{eqn:Tyy_full}
  \end{equation}

  Let us  first analyse the solution of  this equation under the
  often-made assumption that $\tyy$ is constant in
  $x$~\cite{phanthien1983} and then analyse why this solution
  misses an important part of the physics. One might naively
  think that this solution can be seen as the first term in a series
  expansion in powers of $x$~\cite{oztekin1997}, but as we shall see
  this assumption itself is incorrect: singular terms are typically
  generated by the stagnation point flow. Defining
  \begin{equation}
    \tyy(x, y) \equiv Y(y),
  \end{equation}
  we solve
  \begin{equation}
    Y(y) + \De\left[y^2 Y'(y) - 4y Y(y)\right]=4y.
    \label{eqn:ucm_nox}
  \end{equation}
  This equation allows an exact solution, and we find
  \begin{equation}
    Y_\textrm{gen} (y) = 4 y + 12 \De y^2 + 24 \De^2 y^3 + 24 \De^3 y^4
         + C y^4 \exp[1/(\De y)],
    \label{eqn:sol_ucm_nox}
  \end{equation}
  where the last term corresponds to the homogeneous solution of the
  differential equation, and $C$ is a constant of integration. This term
  clearly leads to unphysical results, as it implies $|Y(y)| \to \infty$
  as $y \to 0^+$. Hence, if one thinks of $Y(y)$ as the first term
  in a regular series expansion in $x$, we should discard it and keep
  only the particular solution
  \begin{equation}
    Y_0(y) = 4 y + 12 \De y^2 + 24 \De^2 y^3 + 24 \De^3 y^4.
    \label{eqn:Y0}
  \end{equation}

  A few remarks about this solution are in order:
  {\em (i)} It does not in general satisfy momentum balance, but it is
  qualitatively similar to the more accurate similarity solution of
  \"Oztekin et al.~\cite{oztekin1997}.
  {\em (ii)} By forcing the solution to be independent of $x$, we find
  a solution that can only be seen as an approximation that is valid
  in a small range around $x=0$.
  {\em (iii)} We assume the same functional form of the velocity field
  for all $\De$.
  {\em (iv)} The solution does not diverge at finite $y$ for any $\De$.
  This is in contrast with purely extensional flow, where stresses
  diverge for $\De \geq 1/2$. 
  
  Let us now discuss the shortcoming of the above line of analysis. The
  solution is an admissible solution if we work in the infinite domain
  $x \to \pm \infty$. However, if we work in a finite domain, with
  boundary conditions at some finite $x = \pm L$, say, then the solution
  is only relevant if the stresses on this boundary are precisely
  consistent with (\ref{eqn:Y0}). In general, the boundary stresses are
  incompatible with this expression and the flow drawn in
  Fig.~\ref{fig:flow_geom} advects all deviations from this expression
  towards the stagnation point. While in the above analysis it appears
  as if the analysis close to the
  stagnation point dictates what the proper boundary condition far to
  the left and right should be, in reality we have to analyse what
  happens with the stresses as they are advected from the left and right
  to the stagnation point. That dictates the behaviour there, not the
  other way around! The analysis below does show that singular behaviour
  is picked up through this advection from the boundaries, as one might
  expect.
  
  \begin{figure}[t]
    \begin{center}
      \includegraphics[width=0.4\textwidth]{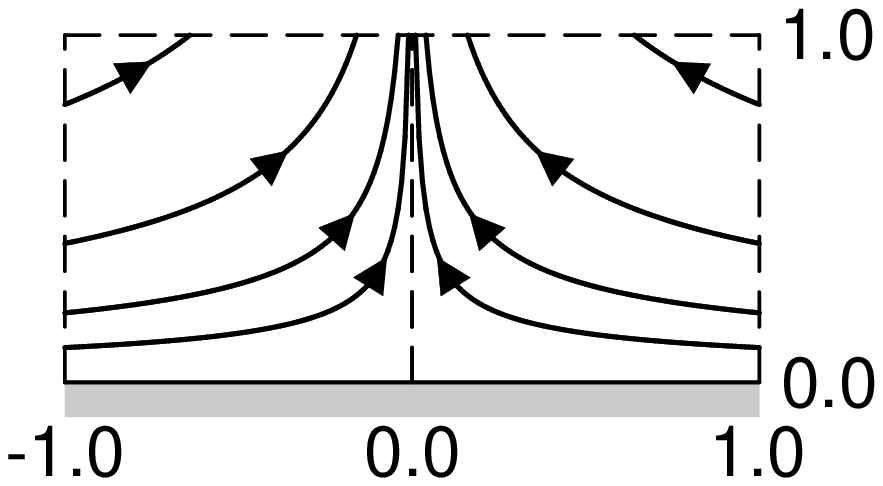}
      \caption{\label{fig:fin_dom} Planar stagnation flow on the  finite
         rectangular domain analysed in this paper.}
    \end{center}
  \end{figure}

  \section{Inflow boundary conditions and an explicit solution}
  To back up the above observations, we now  allow explicit
  $x$-dependence of the stress field  \emph{without} including higher
  orders in $x$ in the velocity field. We can then match more realistic
  inflow boundary conditions by including homogeneous parts of the
  solution, and we shall see that this causes qualitative changes in
  the stress field.

  To do so, we modify the geometry by explicitly taking a finite domain
  in the $x$ direction, with $-1 \leq x \leq 1$, see
  Fig.~\ref{fig:fin_dom}. We observe that the equation is purely
  advective. This implies that is it a first-order differential
  equation and that there is no ``interaction'' across streamlines.
  We can therefore split the domain in two separate parts, one for
  $x < 0$ and one for $x > 0$, with the line $x = 0$ acting as the
  separatrix between the two domains. As in the case of pure extensional
  flow, this allows for non-analyticity on the line
  $x = 0$.~\cite{renardy2006,thomases2006,becherer2008}. On each of
  the two subdomains, we can impose independent inflow boundary
  conditions. We will restrict ourselves to the domain
  $0 \leq x \leq 1$, and we will assume that the negative domain is
  the mirror image of this domain. On the lateral boundary $x = +1$ we
  impose the boundary condition $T_{yy}(x=1,y)=Y_\textrm{b}(y)$ with
  $Y_\textrm{b}(0) = 0$.

  We thus consider again the differential
  equation~(\ref{eqn:Tyy_full}):
  \begin{equation}
    Y(x,y) + \De \left[ -2 x y \frac{\pd Y(x,y)}{\pd x}
                      + y^2  \frac{\pd Y(x,y)}{\pd y}
                      - 4 y Y(x,y)
               \right] = 4 y.
    \label{eqn:Y_full}
  \end{equation}
  In the language of local analysis of differential equations, this
  partial differential equation has an \emph{irregular} singular point
  at $x=y=0$~\cite{benderorszag1978}.

  It is easy to check that the partial differential terms in the
  equation together give identically zero if $Y$ is a function of $xy^2$
  only. In other words, the form
  \begin{equation}
    Y(x,y) \equiv f(xy^2),
  \end{equation}
  is a zero mode of the differential operator since
  \begin{equation}
    -2 x y \frac{\pd f(xy^2)}{\pd x} + y^2 \frac{\pd f(xy^2)}{\pd y}
    \equiv 0
    \label{eqn:advect_part}
  \end{equation}
  for {\em any} function $f$.
  
  Now, due to the linearity of the equation, the solution of the
  equation is the sum of the particular solution $Y_0$ given
  in~(\ref{eqn:Y0}) plus an arbitrary solution $Y_\textrm{hom}(x,y)$ of
  the homogeneous equation, that is
  \begin{equation}
    Y(x,y) = Y_0(y) + Y_\textrm{hom}(x,y).
  \label{eq:y0+hom}
  \end{equation}
  Moreover, if we assume $Y_\textrm{hom} $ to be of the form
  \begin{equation}
    Y_\textrm{hom} (x,y) = g(y) h(xy^2),
  \end{equation}
  then
  inserting this into the homogeneous part of the equation (with
  right hand side equal to zero) effectively gives an equation for
  $g(y)$ that is identical to the homogeneous part of
  Eq.~(\ref{eqn:ucm_nox}). For $g$ we thus recover the homogeneous
  solution of (\ref{eqn:ucm_nox}), that is, the last term
  of~(\ref{eqn:sol_ucm_nox}). Thus, we have
   \begin{equation}
    Y (x,y) = Y_0(y) +   y^4 \exp[1/(\De y)] h(xy^2)  
    \label{eqn:general}
  \end{equation}
  
  We can now impose the boundary condition $Y(x=1,y)=Y_\textrm{b}(y)$ at
  $x=1$: requiring that $Y(x=1,y)$ given by (\ref{eqn:general}) equals
  $Y_\textrm{b} (y)$, immediately gives
      \begin{equation}
    h( y^2) =
        \frac{Y_\textrm{b} (y)
              - Y_0(y)}
              {y^4}\,  \exp[-1/(\De y)].
    \label{eqn:heq0}
  \end{equation}
  Since $h(xy^2)$ is a function of the product $xy^2$ only, we now know
  $h$ for all $x$ and $y$. This finally gives the general full solution
    \begin{equation}
      Y(x,y) = Y_0(y) + 
          \frac{Y_\textrm{b}\left(\sqrt{x y^2}\right)
                - Y_0\left(\sqrt{x y^2}\right)}%
                {x^2} \,  \exp\left[\frac{1-1/\sqrt{x}}{\De y}\right] .
      \label{eqn:heq}
    \end{equation}
  Note that the structure is precisely as we envisioned: the deviations
  of the $x$-independent solution $Y_0 (y)$ from the stress boundary
  condition $Y_\textrm{b}(y)$ are convected towards the stagnation point,
  and the stresses there are largely dominated by the singularity. 

\begin{figure}
    \setlength{\tabcolsep}{0mm}
    \begin{center}
      \begin{tabular}{ccc}
        \includegraphics[height=0.18\textwidth]{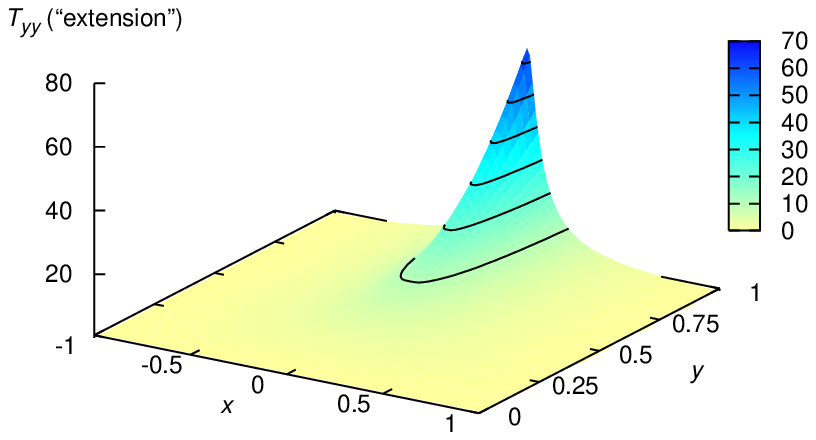} &
        \includegraphics[height=0.18\textwidth]{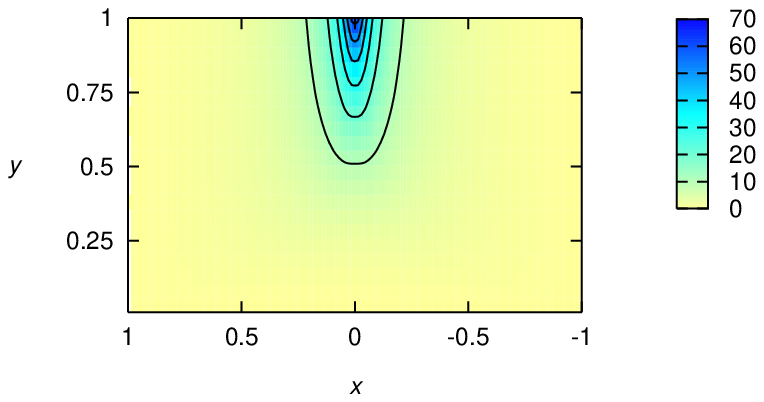} &
        \includegraphics[height=0.18\textwidth]{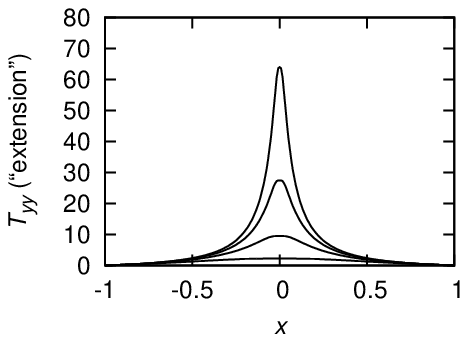} \\
        (a) & (b) & (c)
      \end{tabular}
    \end{center}
    \caption{\label{fig:results}Polymer extension in a simplified
    wall stagnation flow for ultradilute UCM fluid, for $\De = 1.0$:
    (a) three-dimensional
    plot of the extension as a function of $x$ and $y$, (b) contour plot
    of the extension, showing the shape of the birefringent strand, (c)
    cross-section of the extension profile for $y = 0.25,0.5,0.75,1.0$}
  \end{figure}

  For the flow that we consider, we use the procedure
  above to match the stress field to the inflow boundary condition.
  For simplicity we take this to be
  \begin{equation}
    Y_b(y) \equiv 0,
  \end{equation}
  but since the behaviour for small $x$ is dominated by the singular
  term, other choices lead to similar conclusions. For $\De = 1$ we
  then find the stress field in Fig.~\ref{fig:results}. Since we
  consider the ultradilute limit, one may also think of this as
  the polymer extension in the $yy$ direction (see below).

  We clearly see that a narrow region of high extensional stress
  (polymer extension) forms in the centre of the domain. The solution
  we find does not diverge for any $\De$, nor does any gradient
  diverge --- in fact, the $x$-derivatives are zero to any order on the
  centre line, and we have an essential singularity there. 

  \section{Distance from the wall}
\begin{figure}
    \setlength{\tabcolsep}{0mm}
    \begin{center}
      \begin{tabular}{ccc}
        \includegraphics[height=0.2\textwidth]{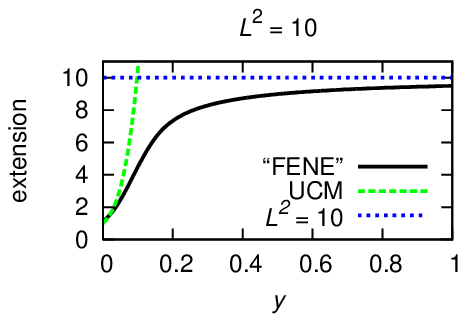} &
        \includegraphics[height=0.2\textwidth]{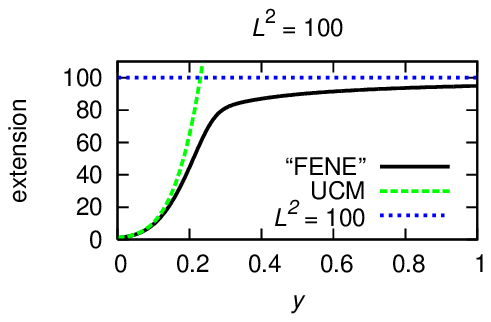} &
        \includegraphics[height=0.2\textwidth]{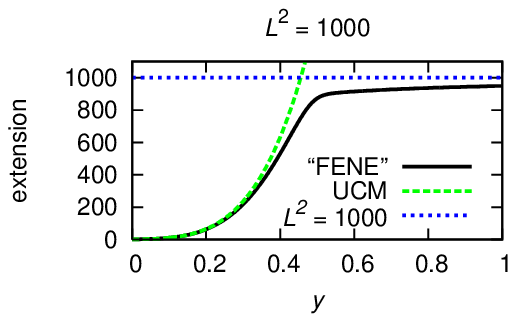} 
      \end{tabular}
    \end{center}
    \caption{\label{fig:comp}Polymer extension in a simplified
    wall stagnation flow for ultradilute FENE-P fluid, for $\De = 5.0$,
    and $L^2 = 10.0, 100.0, 1000.0$.}
  \end{figure}

  The stress profile $Y_0(y)$, Eq.~(\ref{eqn:Y0}), for a UCM fluid
  has no intrinsic length scale, as the UCM model has no characteristic
  stress or extension. Therefore, we cannot obtain a meaningful
  ``distance from the wall'' in that model. Models with finite
  extensibility, such as the well-known FENE-P model, are required to
  do this. In previous work we argued~\cite{becherer2008} that for
  extensional flow a fair approximation to the FENE-P model can be
  obtained by restricting to the $yy$ extensional
  component of the stress tensor, and making the simple approximation
  that the polymer stretching follows the UCM rheology up to a certain
  maximum extensional stress after which the stress does not increase
  any further (this is sometimes referred to as a linear-locked
  approximation~\cite{harris1993,harris1994}). This maximal stress
  should correspond to the maximal extension $L$ in the FENE-P model,
  for which the trace of the conformation tensor $\tens{A}$ is $L^2$.
  The UCM stress tensor is related to the FENE-P conformation
  tensor as
  $\tens{A} = \tens{1} + \De \tens{T}$ (this can be seen by taking
  $L \to \infty$ in the FENE-P constitutive equations, and comparing
  this to the UCM constitutive equations). Restricting to the
  $yy$ component, we find $1 + \De T_{yy}^\textrm{max} = L^2$. In
  Fig.~\ref{fig:comp} we show stress profiles for the simplified
  velocity field, for the UCM model and a FENE-P-like model that is
  restricted to the $yy$ component.

  In this approximation and for a fixed velocity field, we
  can thus estimate the distance of the strand from the wall, as well
  as the width of the strand, by considering the locus of points where
  the UCM extension $1 + \De \tyy$ reaches $L^2$.  For our fixed
  velocity field, it is straightforward to find scaling relations for
  the distance and the width as functions of $\De$ and $L^2$. From
  Eq.~(\ref{eqn:Y0}) we see that the distance $y_0$ from the wall is
  given in our approximation by
  \begin{equation}
    1 + \De y_0 + 12 \De^2 y_0^2 + 24 \De^3 y_0^3 + 24 \De^4 y_0^4
    = L^2.
  \end{equation}
  This means that $y_0$ scales as $1/\De$ for fixed $L^2 \gg 1$ and
  \emph{all} $\De$. For fixed $\De$ we find that $y_0$ should scale as
  $(L^2)^{1/4} = \sqrt{L}$ for $L^2 \to \infty$. We conclude that
  \begin{equation}
    y_0 = \frac{\sqrt{L}}{\De} \frac{1}{24^{1/4}} f(L^2) \quad \textrm{with}
    \quad f(L^2) < 1 \textrm{ and } f(L^2 \to \infty) = 1.
  \end{equation}
  This asymptotic result cannot be easily extended to other, more
  realistic velocity fields. Note that in the limit of large $L^2$,
  the distance $y_0$ will become larger than $1$, and it falls outside
  the domain that we assumed, $0 \leq y \leq 1$. The general conclusion
  that the distance increases with $L^2$ and decreases with $\De$ is of
  course physically reasonable: for larger $L^2$ at fixed $\De$ it
  takes a longer distance to fully stretch the polymers, while for
  larger $\De$ at given $L^2$, the stretching occurs more rapidly and
  one may expect that the distance required for full stretching
  decreases, at least on the centre line of the flow.

\section{Width of the strand}
\begin{figure}
    \begin{center}
      \begin{tabular}{cc}
        \includegraphics[width=0.4\textwidth]{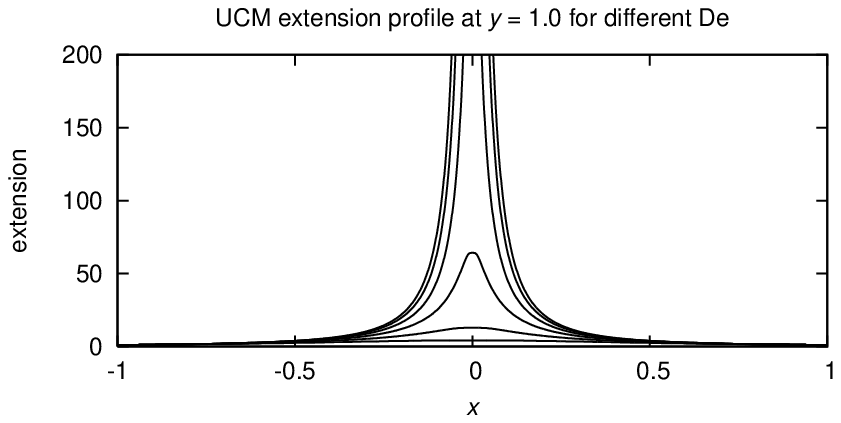} &
        \includegraphics[width=0.4\textwidth]{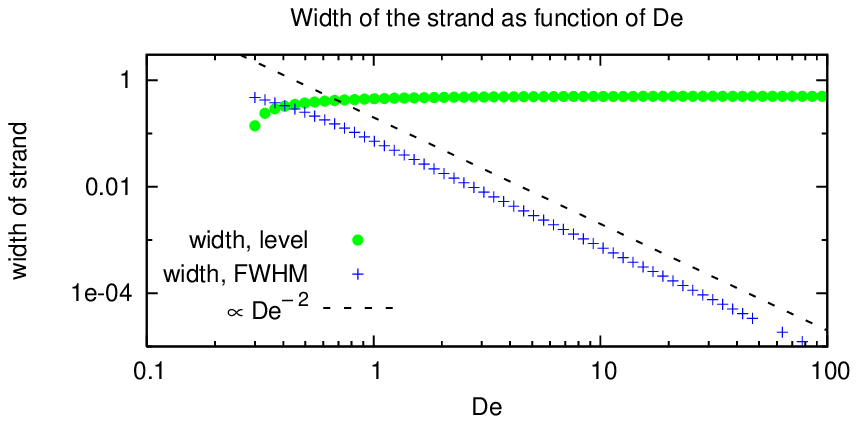} 
      \end{tabular}
    \end{center}
    \caption{\label{fig:width}The left panel shows the cross-section
    of the ``UCM extension'' profile at $y = 1.0$ for a range of
    Deborah numbers, $\De = 0.30, 0.55, 1.00, 1.81, 3.31, 6.03$. The
    right panel shows a log-log plot of the width of the peak, both
    as a FWHM measure and for a FENE-P-like definition (see text).}
  \end{figure}

  The width of the strand is a more subtle issue. Since the stress
  is bounded, and therefore also the UCM extension, we can define
  a meaningful length scale even in the UCM model, as can be seen from
  the left panel of Fig.~\ref{fig:width}. For example, we can take the
  \emph{full width at half maximum} (FWHM) of the peak.
  In that case, the width decreases as a function of $\De$. We can
  also look at our FENE-P-like approximation and use the contour line
  where the profile reaches a given $L^2$. A numerical analysis shows
  that the FWHM decreases as $1 / \De^2$, while the approximate FENE-P
  width increases monotonically and saturates, see the right panel
  of Fig.~\ref{fig:width}.

  These asymptotic results can be explained from the exact
  solution, Eq.~(\ref{eqn:heq}). The FWHM is mostly determined by the
  behaviour close to the centre of the strand, where the factor
  $\exp[-1/(\De\sqrt{x})]$ dominates: as $x \to 0$ it vanishes more
  rapidly than the factors in front of it diverge. The typical length
  scale $x_0$ for this exponential factor is given by
  $\sqrt{x_0} \De \sim 1$, or $x_0 \propto 1/\De^2$, and thus, the
  FWHM also scales as $1/\De^2$. For the FENE-P-like width we make the
  observation that the UCM extension as a function of $x$ actually
  converges pointwise as $\De \to \infty$:
  \begin{equation}
    \lim_{\De \to \infty} \left(1 + \De Y(x, y)\right)
    = \frac{1}{x^2} \quad \textrm{(pointwise in $x$).}
  \end{equation}
  It is then evident that for given $L^2$, the width will approach
  a constant value of $2 / L^2$ for $\De \to \infty$.
  
  \section{Discussion}
  One may ask the question to what extent the analysis above is
  relevant for ``real'' flow profiles. We already argued above that
  although the details may (and will) change, the qualitative behaviour
  (localization of the extensional stress along the separatrix and
  non-analyticity) will persist.

  The procedure that was performed to obtain the stress profile can be
  repeated for the ``exact'' 
  velocity field in Refs.~\cite{phanthien1983} and~\cite{oztekin1997},
  using numerical integration where necessary.
  That would still yield a solution that does not obey the momentum
  balance equation~(\ref{eqn:ns}), but more importantly, it would
  not qualitatively change the singularity that we find.
  Mathematically, this happens as follows. In the region near the
  stagnation point $x=y=0$, the velocity field is expected to have, in
  lowest order, always terms like in (\ref{eq:veq}). Hence, in this
  regime, the equation always has ``zero mode'' solutions $f(xy^2)$ of
  the convective part of the constitutive equation. Only in special
  cases will this term not be excited. Furthermore, close to the wall,
  the dominant behaviour of the $y$-independent homogeneous solution
  must go as $\exp[-1/(\De y)]$ and this then has to be compensated by
  a similar essential singularity in $\sqrt{x} y$. Thus, close to the
  wall this essentially singular behaviour is robust. Further away
  from the wall, however, the behaviour will change if the velocity
  field crosses over to a different $y$-dependence.

  Similar considerations apply to
  other modifications that are used to give a more realistic description
  within the UCM framework. As long as the velocity field is reasonably
  well-behaved, the extension profile on the separatrix is determined
  only by the velocity on the separatrix.

  In fact, we see that the only ingredients that we need in order to
  obtain our results, are the general form of the velocity field 
  Eq.~(\ref{eqn:simsol}), the ``advective'' part of the constitutive
  equation, as in Eq.~(\ref{eqn:advect_part}), and the absence of a
  stress diffusion term in the constitutive equation. This strongly
  suggests
  that both the localization and the nonanalytic behaviour at $x = 0$
  depend on these ingredients and not so much on the particular
  form of the constitutive equations. One would then expect that
  every viscoelastic flow with a stagnation point gives rise to
  singular behaviour downstream of the stagnation point.

  We already mentioned that the stress profile we obtained does not
  satisfy the Navier-Stokes equation (momentum balance). It would be
  interesting to see how we might modify the velocity profile to
  have the system obey the Navier-Stokes equation. It is difficult to
  do this analytically, but given the singular structure of the
  stress field, it is to be expected that a correction to the velocity
  field will show a similar singularity for intermediate flow rates.
  As we mentioned in the introduction, for sufficiently large flow
  rates, this coupling may cause a breakdown of the steady state
  flow solution~\cite{fan2005,hulsen2005}.
  
  \section{Conclusions}
  Not only purely extensional flow but also (reverse) wall stagnation
  flow shows strong localization of extensional stresses and
  non-analytic behaviour. At moderate and high Deborah numers, a
  birefringent strand is
  formed, with an essential singularity at its centre. The polymer
  extension becomes appreciable at a finite distance from the wall, and
  the width of the strand either decreases or becomes constant
  with increasing Deborah number, depending on the definition.
  Asymptotic scaling relations can be found for both the distance from
  the wall and the strand width. Although concrete numerical and
  analytical results have been calculated only for a fixed velocity
  field, the qualitative behaviour that we find should carry over to
  more realistic flows, at least if the usual constitutive equations
  are to be taken seriously.  If an artificial stress diffusion term
  is added to the constitutive equation, as is sometimes done in
  numerical simulations to enhance stability, the mathematical
  singularity will disappear. Nevertheless, the behaviour we have
  analysed here should still dominate most of the profiles if the
  diffusion constant takes on any value that is physically reasonable.

  We suggest that the localization and the singular behaviour are caused
  by the purely advective character (without diffusion) of the
  constitutive equation, combined with the extensional character of
  the flow. We further suggest that the occurrence of this behaviour does
  not depend on the exact formulation of the constitutive equations,
  as long as these include stress advection and stress relaxation. The
  non-analyticity at the separatrix is a result of the absence of
  diffusion, which essentially decouples the domains on either side
  of the separatrix.

  As we noted in the introduction, internal stagnation point
  flows were recently shown to exhibit singular behaviour of the type
  $|y|^\beta$, where $y$ is the upstream distance from the singular
  line. Such singularities may easily cause numerical problems since
  derivatives of sufficiently high order diverge. In the case of wall
  stagnation flow, the situation may at first sight not be as bad for
  numerical approaches, since all derivatives remain finite.
  Nevertheless, for large Deborah numbers there is a rapid crossover
  region, the width of which decreases as $1/\De^2$ for large $\De$.
  This region will have to be resolved numerically.\footnote{%
  Note also that near an essential singularity $f(x) \approx \e^{-1/x}$,
  the ratio $f^{(n)}/f$ diverges as $1/x^{2n}$ for $x \to 0$, so
  finite difference approximations become increasingly bad near an
  essential singularity.}

  \section*{Acknowledgements}
  The authors would like to thank Gareth H. McKinley for bringing this
  subject to their attention. ANM acknowledges support from the
  Royal Society of Edinburgh / BP Trust Research Fellowship

  \bibliography{singular}

\begin{thebibliography}{10}
\expandafter\ifx\csname url\endcsname\relax
  \def\url#1{\texttt{#1}}\fi
\expandafter\ifx\csname urlprefix\endcsname\relax\def\urlprefix{URL }\fi

\bibitem{bird1}
R.~B. Bird, C.~F. Curtiss, R.~C. Armstrong, O.~Hassager, Dynamics of polymeric
  liquids, 2nd Edition, Vol. 1. Fluid Mechanics, Wiley, New York, 1987.

\bibitem{bird2}
R.~B. Bird, C.~F. Curtiss, R.~C. Armstrong, O.~Hassager, Dynamics of polymeric
  liquids, 2nd Edition, Vol. 2. Kinetic theory, Wiley, New York, 1987.

\bibitem{xi2007}
L.~Xi, M.~D. Graham, A mechanism for oscillatory instability in viscoelastic
  cross-slot flow, J. Fluid Mech., in press. \url{arXiv:physics/0703262}
  (2007).

\bibitem{arratia2006}
P.~E. Arratia, C.~C. Thomas, J.~Diorio, J.~P. Gollub, Elastic instabilities of
  polymer solutions in cross-channel flow, Phys. Rev. Lett. 96 (2006) 144502.

\bibitem{poole2007}
R.~J. Poole, M.~A. Alves, P.~J. Oliveira, Purely elastic flow asymmetries,
  Phys. Rev. Lett. 99 (2007) 164503.

\bibitem{renardy2006}
M.~Renardy, A comment on smoothness of viscoelastic stresses, J. Non-Newtonian
  Fluid Mech. 138 (2006) 204--205.

\bibitem{thomases2006}
B.~Thomases, M.~Shelley, Emergence of singular structures in {O}ldroyd-{B}
  fluids, Phys. Fluids 19 (2007) 103103.

\bibitem{becherer2008}
P.~Becherer, A.~N. Morozov, W.~van Saarloos, Scaling of singular structures in
  extensional flow of dilute polymer solutions, J. Non-Newtonian Fluid Mech.
  153 (2008) 183--190.

\bibitem{larson}
R.~G. Larson, The Structure and Rheology of Complex Fluids, Oxford University
  Press, Oxford, 1999.

\bibitem{benderorszag1978}
C.~M. Bender, S.~A. Orszag, Advanced mathematical methods for scientists and
  engineers, McGraw-Hill, New York, 1978.

\bibitem{wapperom2005}
P.~Wapperom, M.~Renardy, Numerical prediction of the boundary layers in the
  flow around a cylinder using a fixed velocity field, J. Non-Newtonian Fluid
  Mech. 125 (2005) 35--48.

\bibitem{fan2005}
Y.~Fan, H.~Yang, R.~I. Tanner, Stress boundary layers in the viscoelastic flow
  past a cylinder in a channel: limiting solutions, Acta Mech. Sinica 21 (2005)
  311--321.

\bibitem{hulsen2005}
M.~A. Hulsen, R.~Fattal, R.~Kupferman, Flow of viscoelastic fluids past a
  cylinder at high {Weissenberg} number: stabilized solutions, J. Non-Newtonian
  Fluid Mech. 127 (2005) 27--39.

\bibitem{cressely1980}
R.~Cressely, R.~Hocquart, Birefringence d'\'ecoulement localis\'ee induite \`a
  l'arri\`ere d'obstacles, Optica Acta 27 (1980) 699--711.

\bibitem{harlen1990b}
O.~G. Harlen, J.~M. Rallison, M.~D. Chilcott, High-{Deborah}-number flow of a
  dilute polymer solution past a sphere falling along the axis of a cylindrical
  tube, J. Non-Newtonian Fluid Mech. 33 (1990) 157--173.

\bibitem{bajaj2008}
M.~Bajaj, M.~Pasquali, J.~Ravi~Prakash, Coil-stretch transition and the
  breakdown of computations for viscoelastic fluid flow around a confined
  cylinder, J. Rheol. 52 (2008) 197--223.

\bibitem{phanthien1983}
N.~Phan-Thien, Plane and axi-symmetric stagnation flow of a {M}axwellian fluid,
  Rheol. Acta 22 (1983) 127--130.

\bibitem{oztekin1997}
A.~\"{O}ztekin, B.~Alakus, G.~H. McKinley, Stability of planar stagnation flow
  of a highly viscoelastic fluid, J. Non-Newtonian Fluid Mech. 72 (1997) 1--29.

\bibitem{harris1993}
O.~J. Harris, J.~M. Rallison, Start-up of a strongly extensional flow of a
  dilute polymer solution, J. Non-Newtonian Fluid Mech. 50 (1993) 89--124.

\bibitem{harris1994}
O.~J. Harris, J.~M. Rallison, Instabilities of a stagnation point flow of a
  dilute polymer solution, J. Non-Newtonian Fluid Mech. 55 (1994) 59--90.

\end{thebibliography}
    \bibliographystyle{elsart-num}

\end{document}